\documentclass[aps,pra,superscriptaddress,showpacs]{revtex4}

\usepackage{amsmath}
\usepackage{amssymb}
\usepackage{graphicx}
\usepackage{color}

\newcommand{\beq}{\begin{equation}}
\newcommand{\eeq}{\end{equation}}
\newcommand{\bea}{\begin{eqnarray}}
\newcommand{\eea}{\end{eqnarray}}
\newcommand{\hf} {\frac{1}{2}}

\def\rm#1{{\mathrm#1}}

\newcommand{\addrHD}{Max--Planck--Institut f\"ur Kernphysik,
Saupfercheckweg 1, 69117 Heidelberg, Germany}

\newcommand{\addrUD}{Department of Theoretical Physics, University of Debrecen,
Debrecen, Hungary}

\newcommand{\addrIN}{Institute of Nuclear Research, P.~O.~Box 51,
Debrecen, Hungary}

\begin{document}

\sloppy

\title{On the applicability of the layered sine--Gordon model\\ 
for Josephson-coupled high-${\boldsymbol T}_{\rm c}$-layered superconductors}

\author{I. N\'andori}
\affiliation{\addrIN}
\affiliation{\addrHD}

\author{U. D. Jentschura}
\affiliation{\addrHD}

\author{S. Nagy}
\affiliation{\addrUD}

\author{K. Sailer}
\affiliation{\addrUD}

\author{K. Vad}
\affiliation{\addrIN}

\author{S. M\'esz\'aros}
\affiliation{\addrIN}

\begin{abstract}
We find a mapping of the layered sine--Gordon model to an equivalent 
gas of topological excitations and determine the long-range interaction
potentials of the topological defects. This enables us to make a
detailed comparison to the so-called layered vortex gas, which can be 
obtained from the layered Ginzburg--Landau model.
The layered sine--Gordon model has been proposed in the literature 
as a candidate field-theoretical model for Josephson--coupled 
high-$T_{\rm c}$ superconductors, and the implications of our 
analysis for the applicability of the layered sine--Gordon model
to high-$T_{\rm c}$ superconductors are discussed.
We are led to the conjecture 
that the layered sine--Gordon and the layered vortex gas models belong to 
different universality classes.
The determination of the critical temperature of the 
layered sine--Gordon model is based on a renormalization-group analysis.
\end{abstract}


\maketitle

%
%
\section{INTRODUCTION}
\label{intro} 

Two essential prerequisites for an analysis of superconducting
materials are anisotropic models, 
as initiated by Ginzburg~\cite{Gi1952}, and the inclusion of vortices,
as envisaged by Abrikosov~\cite{Ab1957}.
Typical high transition temperature superconductors consist of 
copper-oxide superconducting planes separated by insulating layers.
In the phenomenological description of high-$T_{\rm c}$ 
superconductivity, one may use an anisotropic, continuous Ginzburg--Landau 
theory~\cite{Gi1952,BlEtAl1994,ChDuGu1995}, 
but only for not too large anisotropy.
Note that the anisotropic, continuous model can be mapped 
onto the isotropic Ginzburg--Landau model by an 
appropriate rescaling method~\cite{RESCALE}. However,
in the case of extremely high anisotropy like in 
$\mbox{Bi}_2 \mbox{Sr}_2 \mbox{Ca} \mbox{Cu}_2 \mbox{O}_8$, the 
discreteness of the structure becomes relevant \cite{Mo2005}, and it becomes 
necessary to use a layered Ginzburg--Landau model~\cite{BlEtAl1994,LaDo1971} 
where the layers are coupled by Josephson or electromagnetic 
interactions. This provides a good basis for the discussion of the 
vortex dominated properties of high-$T_{\rm c}$ superconductors. 
Some exact and some approximate mappings of the layered Ginzburg--Landau 
model (i.e.~the Lawrence-Doniach model~\cite{LaDo1971})
onto various other statistical, field theoretical or spin models, 
like the layered vortex gas \cite{Pi1994,Pi1995prl1,Pi1995prl2,%
Pi1995prb,Pi1997pm,BlEtAl1994} and the anisotropic 
XY~models~\cite{3DXY,HiTs1980,Kh2006} have already been discussed 
in the literature, and these models 
have also been proposed and used for the description of the vortex 
dynamics in high-$T_{\rm c}$ superconductors. 
Connections to sine--Gordon type models 
\cite{PiVa1992,PiVaBa1992,PiVa1994,Ro2000,NaNaSaJe2005,NaSa2006,JeNaZJ2006,BeCaGi2007} 
have also been explored in the literature.

The latter investigations are motivated by the well-known fact that the 
massless two-dimensional
(2D) sine--Gordon scalar field theory belongs to the universality 
class of the 2D--XY spin model and consequently to that of the 2D Coulomb 
or vortex gas. The mappings between these models 
and also the phase structure have been discussed 
in the literature in great detail (see, e.g., Refs.~\cite{NaPoSa2001,%
NaPoSa2001pmb,Pi1994,Pi1995prl1,Pi1995prl2,Pi1995prb,Pi1997pm,PiVa1992,%
PiVaBa1992,PiVa1994,ZJ1996,SGModel,ThirringModel,Co1975}). 
Since the layered Ginzburg--Landau model 
can be considered as the continuum limit of the anisotropic 3D--XY model 
(discrete in the $z$-direction but continuous in the $xy$-planes), one 
might suggest that the field theoretical counterpart of the layered 
Ginzburg--Landau model should be a sine--Gordon type model. 
However, the 3D Ginzburg--Landau theory (in the London limit and 
in the absence 
of electromagnetic fields), which can be considered as the continuum limit 
of the 3D--XY planar rotator, and the 3D sine--Gordon model do not belong 
to the same universality class 
(see~Refs.~\cite{NaJeSaSo2004,NaSaJeSo2002,Ko1977}), 
a phase transition being absent in the 3D sine--Gordon case. Since
layered models are always constructed from 3D models by a suitable 
discretization of the derivative term in one of the spatial dimensions 
(see, e.g., Ref.~\cite{Na2006}), the equivalence of the layered vortex 
gas and layered sine--Gordon models remains questionable. One purpose 
of this paper is to clarify this point by finding an exact mapping of the 
layered sine--Gordon model to an equivalent gas of topological excitations, 
which in turn can be compared directly to the layered vortex gas. 
We also consider the phase 
structure and the critical behaviour of 
the $N$-layer sine--Gordon model by a renormalization group method, 
and determine the relation of the critical parameter $b_{\rm c}^2$
of the layered sine--Gordon model to the 
critical temperature, as a function of the number of 
layers.

This paper is organized as follows.
In Sec.~\ref{lsgsect}, we discuss the comparison of the 
layered Ginzburg--Landau to the layered sine--Gordon model,
by a mapping of each model to an equivalent gas of topological 
excitations.  In Sec.~\ref{rg}, we discuss the 
the renormalization-group
(RG) flow of the layered sine--Gordon
model. Conclusions are reserved for Sec.~\ref{sum}.

%
%
\section{LAYERED GINZBURG--LANDAU VERSUS LAYERED SINE--GORDON MODEL}
\label{lsgsect}

%
%
\subsection{Mapping of the layered Ginzburg--Landau model to the 
layered vortex gas}

The Ginzburg--Landau theory has been developed by applying a variational 
method to an assumed expansion of the free energy in powers of 
$\vert\psi\vert^2$ and $\vert\partial_{\mu}\psi\vert^2$ where $\psi$ 
is a complex order parameter (the inhomogeneous condensate of the 
superconducting electron pairs) and $\vert\psi\vert^2$ represents the 
local density of superconducting electron pairs 
(for a detailed discussion see, e.g., Ref.~\cite{Le1997}). 
Its detailed form can be found in Eqs.~(6--6) and (6--9) of 
Ref.~\cite{dG1999}.
Upon a discretization of one of the spatial directions
(say, the $z$-coordinate), one obtains the layered Ginzburg--Landau
(or Lawrence--Doniach~\cite{LaDo1971}) model with the free energy
(in natural units: $\hbar = c = \epsilon_0 = 1$),
\begin{equation}
\label{LGL}
F = s \int {\rm d}^2 r \, 
\left(\sum_{n=1}^N\left(\alpha \vert \psi_n \vert^2 + 
{\beta\over 2} \vert \psi_n \vert^4  
+ \frac{\vert \partial_x \psi_n \vert^2 + 
\vert \partial_y \psi_n \vert^2}{2 \, m_{ab}}\right)
+ \sum_{n=1}^{N-1}
{\vert \psi_{n+1} - \psi_{n} \vert^2\over 2 \, m_{c} \, s^2} \right)\,.
\end{equation}
Here, $m_{ab}$ and $m_{c}$ represent the intralayer and interlayer effective 
masses, $s$ is the interlayer distance, and $N$ stands for the total number 
of layers. The sum over $\mu$ covers the spatial coordinates 
$\mu=x,y,z$. The parameters $\alpha$ and $\beta$ are 
discussed in Eqs.~(6--6) and (6--8) of Ref.~\cite{dG1999}.
In order to investigate the vortex dynamics in the framework of 
the  Ginzburg--Landau theory, one has to consider the discretized model
given in Eq.~(\ref{LGL}) in the London approximation. 
One writes the complex, layer-dependent order parameter as 
$\psi_n(\vec{r}) = \psi_{0,n}(\vec{r}) \, \exp[{\rm i}\phi_{n}(\vec{r})]$
with real $\psi_{0,n}(r)$, where the $\phi_{n} \in [0,2\pi)$ 
are compact variables, 
and the moduli $\psi_{0,n}$ are assumed to be constant and identical in 
every layer (i.e. $\psi_{0,n}(\vec{r}) = \psi_{0}$) which is the London-type 
approximation. The London-type form of the layered Ginzburg--Landau 
model with Josephson coupling can be mapped 
(see Refs.~\cite{Pi1994,Pi1995prl1,Pi1995prl2,Pi1995prb,Pi1997pm,BlEtAl1994}) 
onto the layered vortex-gas. 
The globally neutral layered vortex-gas with $N$ layers is characterized 
by the partition function (see Eq.~(2.3) of Ref.~\cite{Pi1995prb})
\begin{equation}
\label{LVG}
{\cal Z}_{\mathrm{LVG}} = \sum_{\nu =0}^{\infty} \frac{z^{2\nu}}{(\nu !)^2} 
\sum_{n_1=1}^N \int \frac{{\rm d}^2 r_1}{a^2} 
\, \dots \,  \sum_{n_{2\nu}=1}^N \int \frac{{\rm d}^2 r_{2\nu}}{a^2} \,
\sum_{\begin{array}{c} 
\scriptstyle \sigma_1, \dots, \sigma_{\nu} = \pm 1 \\
\scriptstyle \sigma_{\nu+\gamma} = -\sigma_\gamma , \;
\gamma \in \{ 1, \dots \nu \} \end{array}}
\exp\left(-\frac{1}{k_{\rm B} T} \, \sum_{\alpha \neq \beta} 
\frac12 \, \sigma_\alpha \, \sigma_\beta \, V(r_{\alpha\beta}, \,n_{\alpha\beta}) 
\right),
\end{equation}
where $\sigma_\alpha = \pm 1$ is the charge of the 
$\alpha$th vortex, $a$ stands for the lattice spacing,
$k_{\rm B}$ is the Boltzmann constant, $T$ is the 
temperature, and the interaction potential $V$ 
between two vortices depends on their relative distance $r_{\alpha\beta}$
within the two-dimensional planes
($r_{\alpha\beta} = \vert \vec{r}_{\alpha} - \vec{r}_{\beta}\vert$) and on 
the distance $n_{\alpha\beta}$ across the planes
($n_{\alpha\beta} = \vert n_\alpha -n_\beta \vert$), 
where $n_\alpha$ is the layer in which the $\alpha$th vortex is 
located. There are $2 \nu$ vortices with fugacity $z$ and these 
fulfill the neutrality condition 
$\sum_{\alpha = 1}^{2 \nu} \sigma_\alpha = 0\,$.
The positive (negative) vorticity is represented by positive 
(negative) charges. 
The restriction to globally neutral charge configurations is ensured
by the condition $\sigma_{\nu+\gamma} = -\sigma_\gamma$ for 
$\gamma \in \{ 1, \dots \nu \}$ in Eq.~(\ref{LVG}).
Following 
Refs.~\cite{Pi1994,Pi1995prl1,Pi1995prl2,Pi1995prb,Pi1997pm,BlEtAl1994}, 
we neglect interactions between vortices separated by more than one layer,
and this results in intra- and interlayer interaction potentials
which have commonly accepted short- and long-range asymptotic forms 
given by (see, e.g., Eqs.~(2.5) and (2.6) of Ref.~\cite{Pi1995prb}) 
\begin{subequations}
\label{int_pot}
\begin{eqnarray}
\label{int_pot_intra}
V(r_{\alpha\beta}, \, n_{\alpha\beta}=0) &=& 
- \ln\left( \frac{r_{\alpha\beta}}{a} \right) -
\sqrt{\lambda} \,\, \frac{r_{\alpha\beta} - a}{a},
\\[2ex]
\label{int_pot_inter}
V(r_{\alpha\beta}, \, n_{\alpha\beta}=1) &=& 
b \, \sqrt{\lambda} \,\, \frac{r_{\alpha\beta}}{a}\,.
\end{eqnarray}
\end{subequations}
The coupling $\lambda \sim a^2 J_{\perp}/J_{\parallel}$ is proportional
to the ratio of the interlayer Josephson coupling $J_{\perp}$ to the 
intralayer coupling $J_{\parallel}$, and $b$ is a constant of order 
unity. The intralayer interaction between the vortices is logarithmic 
for short distances, as in the case of the usual 2D Coulomb or vortex gas, 
but linear for large distances. The interlayer interaction is always 
linear and similar to the long-range intralayer interaction but with an 
opposite sign. Within a layer, vortices of opposite charge attract, whereas 
the positive prefactor of the linear term in the interlayer interaction
implies the formation of vortex stacks of like charges.

%
%
\subsection{Mapping of the layered sine--Gordon model 
to an equivalent gas of topological excitations}

The well-known sine--Gordon model in Euclidean space is defined via the 
action 
\begin{equation}
\label{sg} 
S_{\rm SG}[\varphi] = 
\int {\rm d}^2 r \, \left( \frac12\, (\partial_{\mu} \varphi)^2 
-  y \cos(b \, \varphi)  \right)\,,
\end{equation} 
where the minus sign of the periodic term is chosen so that the zero-field 
configuration remains a (local, not global, infinitely degenerate) minimum. 
As usual, $\varphi$ here is a dimensionless scalar field, 
$(\partial_{\mu}\varphi)^2 \equiv\sum_{\mu=1}^2 (\partial_{\mu}\varphi)^2$, 
$y$ is a fundamental Fourier amplitude, and $b$ is a dimensionless frequency. 
This model is well known to describe the Kosterlitz--Thouless--Berezinskii
(KTB) phase transition~\cite{KTBPhase} in two dimensions. If one adds a 
second layer which leads to the appearance of two fields $\varphi_1$ and 
$\varphi_2$, one may devise the following natural ansatz for the interlayer 
interaction term ${1\over 2}  J (\varphi_1 -\varphi_2)^2$,
where $J$ is the Josephson-type 
coupling whose physical dimension is equal to the square of the inverse 
length. Indeed, the layered sine--Gordon model with this particular 
interlayer interaction term has been proposed in \cite{PiVa1992,PiVaBa1992} 
for the description of the vortex properties of Josephson coupled layered 
superconductors.  The double-layer sine--Gordon 
model~\cite{FiKoSu1979,HeHoIs1995,NaNaSaJe2005,PiVa1992,PiVaBa1992,PiVa1994}
is thus given by the Euclidean action 
\begin{align}
\label{2LSG}
S_{\mathrm{2LSG}} =& \; \int {\rm d}^2 r \, \left(
{1\over 2} \sum_{n=1}^{2}  (\partial_{\mu} \varphi_n)^2 
+  {1\over 2}  J (\varphi_1 -\varphi_2)^2 
-  y \sum_{n=1}^2 \cos(b\varphi_n)  
\right)
\nonumber\\[2ex]
=& \; \int {\rm d}^2 r \, \left(
\frac{1}{2} 
(\partial \underline\varphi)^{\rm T} (\partial \underline\varphi) + 
\frac{1}{2} \,
\underline\varphi^{\rm T} \, {\underline{\underline m}}^2  \underline\varphi
- y \sum_{n=1}^{2} \cos(b \underline{f}_n^{\rm T} \, \underline{\varphi}) 
\right) \,,
\end{align}
where ${\underline \varphi}$ denotes the column vector 
${\underline \varphi} = (\varphi_1, \varphi_2)$ characterizing the 
$O(2)$-doublet, and the $\underline{f}_n$ are projectors
$\underline{f}_n = (\delta_{1n}, \delta_{2n})$ whose components
are given by Kronecker-deltas. 
The Josephson-type interlayer interaction corresponds
to the following dimensionful mass matrix 
(see, e.g., Ref.~\cite{NaNaSaJe2005}),
\begin{equation}
\label{2lsg_mass_matrix}
{\underline{\underline m}}^2 = 
\left(\begin{array}{cc} 
J  & \quad -J  \\[2ex]
-J & \quad J  
\end{array} \right) \,.
\end{equation}
The mass-eigenvalues are $0$ and $2J$. In order to perform the mapping 
of the double-layer sine--Gordon model (\ref{2LSG}) onto a gas of 
topological excitations, we follow the scenario of Ref.~\cite{NaJeSaSo2004}, 
where the partition function of the sine--Gordon model is identically 
rewritten in the form of the partition function of a Coulomb gas. We 
should perhaps note that this mapping procedure is inspired by the treatment 
in Chap.~31 of~\cite{ZJ1996}. One expands the exponential 
factor of the integrand with the periodic potential in a Taylor series, 
expresses $\cos(b \underline{f}_n^{\rm T} 
\underline{\varphi})$ in terms of exponential functions and 
introduces integer valued variables, the charges $\sigma_n=\pm 1$, that 
fulfill the neutrality condition. After these operations, one obtains
\begin{eqnarray}
\label{step1}
{\cal Z}_{\mathrm{2LSG}} 
& = & 
{\mathcal N} \int [{\mathcal D} \underline{\varphi}] 
\exp\left(-S_{\mathrm{2LSG}}[\underline \varphi] \right)
\, = \, 
{\mathcal N} 
\sum_{\nu = 0}^\infty \frac{(y/2)^{2\nu}}{(2\nu)!} 
\sum_{n_{1}=1}^2   \int {\rm d}^2 r_1  \ldots 
\sum_{n_{2\nu}=1}^2 \int {\rm d}^2 r_{2\nu} \,\,
\nonumber \\
&& \times
\sum_{\begin{array}{c} 
\scriptstyle \sigma_1, \dots, \sigma_{\nu} = \pm 1 \\
\scriptstyle \sigma_{\nu+\gamma} = -\sigma_\gamma , \;
\gamma \in \{ 1, \dots \nu \} \end{array}}
\int [{\mathcal D}\underline{\varphi}]
\exp{\left(-\int {\rm d}^2 r \, \left( \frac{1}{2} 
\underline\varphi^{\rm T} \,\,
(-{\underline{\underline 1}}\,\,\partial_{\mu}\partial^{\mu} 
+ {\underline{\underline m}}^2)  
\underline\varphi
+ {\rm i} \, b \, {\underline\rho}^{\rm T} \, \underline\varphi \right)
\right)},
\end{eqnarray}
where $\underline{\underline 1}$ stands for the two-dimensional 
unit-matrix which will be suppressed in the following. The charge density 
${\underline \rho}(\vec{r})$, which depends on the configuration of the 
charges $\sigma_{1}, \ldots, \sigma_{2 \nu}$ and on their positions 
$\vec{r}_1,\ldots,\vec{r}_{2\nu}$, constitutes a vector in the internal 
space of the fields $(\varphi_1, \varphi_2)$ characterizing the two layers 
and reads, ${\underline \rho}(\vec{r}) \equiv \sum_{\alpha =1}^{2\nu} 
\sigma_{\alpha} \, \delta(\vec{r} - \vec{r}_{\alpha}) \,
\underline{f}_{n_{\alpha}}$.
We have thus obtained a representation in which the $2\nu$ charges have been 
placed onto the two layers, with the $\alpha$th charge on layer $n_\alpha$.
Performing the Gaussian path integral in Eq.~(\ref{step1}),
we obtain
\begin{eqnarray}
\label{step2}
{\cal Z}_{\mathrm{2LSG}} & = & {\mathcal N}\, 
\sum_{\nu =0}^\infty \frac{(y/2)^{2\nu}}{(2\nu)!} 
\sum_{n_{1}=1}^2   \int {\rm d}^2 r_1  \ldots 
\sum_{n_{2\nu}=1}^2 \int {\rm d}^2 r_{2\nu} 
\sum_{\begin{array}{c} 
\scriptstyle \sigma_1, \dots, \sigma_{\nu} = \pm 1 \\
\scriptstyle \sigma_{\nu+\gamma} = -\sigma_\gamma , \;
\gamma \in \{ 1, \dots \nu \} \end{array}}
\nonumber \\
&& \times
\exp\left(-\frac{b^2}{2} \int \frac{{\rm d}^2 p}{(2\pi)^2} \,\,
{\underline \rho}^{\rm T}(-\vec{p}) \,
(\vec{p}^{\,2} + {\underline{\underline m}}^2)^{-1} \,  
{\underline \rho}(\vec{p}) 
\right) \,,
\end{eqnarray}
where ${\underline \rho}(\vec{p}) = \sum_{\alpha =1}^{2\nu} 
\sigma_{\alpha} \exp({\rm i} \vec{p}\cdot \vec{r}_{\alpha}) \,
\underline{f}_{n_{\alpha}}$ is the Fourier transform of the 
$O(2)$-charge density. In momentum space, the propagator can easily 
be calculated by matrix inversion, and this gives
\begin{eqnarray}
\label{step4}
{\cal Z}_{\mathrm{2LSG}} & = & {\mathcal N}\, 
\sum_{\nu =0}^\infty \frac{(y/2)^{2\nu}}{(2\nu)!} 
\sum_{n_{1}=1}^2   \int {\rm d}^2 r_1  \ldots 
\sum_{n_{2\nu}=1}^2 \int {\rm d}^2 r_{2\nu} 
\sum_{\begin{array}{c} 
\scriptstyle \sigma_1, \dots, \sigma_{\nu} = \pm 1 \\
\scriptstyle \sigma_{\nu+\gamma} = -\sigma_\gamma , \;
\gamma \in \{ 1, \dots \nu \} \end{array}}
\nonumber \\
&& \times
\exp\left(-b^2
\sum_{\alpha=1}^{2\nu} \, 
\sum_{\gamma=1}^{2\nu} \, 
\frac12 \,
\sigma_{\alpha} \, \sigma_{\gamma}
\left(\begin{array}{cc} 
\delta_{1n_{\alpha}} &\hspace*{0.2cm} \delta_{2n_{\alpha}}
\end{array} \right) \,
\left(\begin{array}{cc} 
A(r_{\alpha \gamma}) &\hspace*{0.5cm}  B(r_{\alpha \gamma})  \\
B(r_{\alpha \gamma}) &\hspace*{0.5cm}  A(r_{\alpha \gamma})
\end{array} \right) \,
\left(\begin{array}{cc} 
\delta_{1n_{\gamma}} \\
\delta_{2n_{\gamma}}  
\end{array} \right)
\right)\,.
\end{eqnarray}
Here, the interaction potentials are
($r_{\alpha\gamma} = | \vec{r}_{\alpha} - \vec{r}_{\gamma} |$)
\begin{subequations}
\begin{align}
\label{intpot_intra}
A(r_{\alpha\gamma}) \; = & \;
\int \, \frac{{\rm d}^2 p}{(2\pi)^2} \,
\frac{e^{[{\rm i} \, \vec{p}\cdot (\vec{r}_{\alpha} - \vec{r}_{\gamma})]} \, 
(\vec{p}^{\,2}+J)}{\vec{p}^{\,2} (\vec{p}^{\,2} +2J)}
= 
-\frac{1}{2\pi} \left(
\frac{1}{2} \ln{\left(\frac{r_{\alpha \gamma}}{a}\right)} -
\frac{1}{2} \left[K_0\left(\frac{r_{\alpha \gamma}}%
{\lambda_{\mathrm{eff}}}\right) 
- K_0\left(\frac{a}{\lambda_{\mathrm{eff}}}\right)\right] 
\right)\,,
\\
\label{intpot_inter}
B(r_{\alpha\gamma}) \; = & \;
\int \, \frac{{\rm d}^2 p}{(2\pi)^2} \,
\frac{e^{[{\rm i} \, \vec{p}\cdot 
(\vec{r}_{\alpha} - \vec{r}_{\gamma})]} \, 
J}{\vec{p}^{\,2} (\vec{p}^{\,2} + 2J)}
=
-\frac{1}{2\pi} \left(
\frac{1}{2}  \ln{\left(\frac{r_{\alpha \gamma}}{a}\right)} +
\frac{1}{2}  \left[K_0\left(\frac{r_{\alpha \gamma}}%
{\lambda_{\mathrm{eff}}}\right) 
- K_0\left(\frac{a}{\lambda_{\mathrm{eff}}}\right)\right] 
\right)\,,
\end{align}
\end{subequations}
where the momentum integrals can be performed using either dimensional 
regularization~\cite{ZJ1996} or ultraviolet (UV) cutoffs and the 
physically relevant, finite parts of the interaction potentials 
consist of massless and massive scalar propagators. In the expression
for the intralayer ($A$) and interlayer ($B$) interaction potentials,
$a$ is the lattice spacing which serves as a short-distance (UV) cutoff,
$K_0$ denotes the modified Bessel function of the second kind, and
$\lambda_{\mathrm{eff}} = 1/\sqrt{2 J}$ is an effective screening length.
The asymptotics of the interaction potentials read as follows
($\gamma_{\rm E} = 0.55721\,56649\dots$ is Euler's constant),
\begin{subequations}
\label{intpot_limits}
\begin{eqnarray}
\label{intpot_intra_short}
A(r_{\alpha \gamma} \ll \lambda_{\mathrm{eff}}) \, &\sim& \, -\frac{1}{2\pi} \,
\ln\left(\frac{r_{\alpha \gamma}}{a}\right),
\\
\label{intpot_intra_long}
A(r_{\alpha \gamma} \gg \lambda_{\mathrm{eff}}) \, &\sim& \, -\frac{1}{2\pi} \,
\left( \frac{1}{2} \, \ln\left(\frac{r_{\alpha \gamma}}{\lambda_{\mathrm{eff}}}\right)
+\, \ln\left(\frac{\lambda_{\mathrm{eff}}}{a}\right) + 
\frac12 \, \ln(2) - \frac12\, \gamma_{\rm E} \right),
\\
\label{intpot_inter_short}
B(r_{\alpha \gamma} \ll \lambda_{\mathrm{eff}}) \, &\sim& \, 0,
\\
\label{intpot_inter_long}
B(r_{\alpha \gamma} \gg \lambda_{\mathrm{eff}}) \, &\sim& \, -\frac{1}{2\pi} \,
\left( \frac{1}{2} \, \ln\left(\frac{r_{\alpha \gamma}}{\lambda_{\mathrm{eff}}}\right)
+ \frac12 \, \ln(2) - \frac12\, \gamma_{\rm E} \right).
\end{eqnarray}
\end{subequations}
Here, $a \ll \lambda_{\mathrm{eff}}$ is assumed. 
The partition function of the double-layer sine--Gordon model is thus 
identically rewritten in the form of a partition function for a gas of 
topological excitations, which we would like to call
the ``layered sine--Gordon gas'' and which is given by 
\begin{eqnarray}
\label{EGTE}
{\cal Z}_{\mathrm{2LSG}} &=& {\mathcal N}\,
\sum_{\nu =0}^\infty \frac{(y/2)^{2\nu}}{(2\nu)!} 
\sum_{n_{1}=1}^2   \int {\rm d}^2 r_1  \ldots 
\sum_{n_{2\nu}=1}^2 \int {\rm d}^2 r_{2\nu} 
\sum_{\begin{array}{c} 
\scriptstyle \sigma_1, \dots, \sigma_{\nu} = \pm 1 \\
\scriptstyle \sigma_{\nu+\gamma} = -\sigma_\gamma , \;
\gamma \in \{ 1, \dots \nu \} \end{array}}
\nonumber \\
&& \times \exp\left(-b^2 \,
\sum_{\alpha \neq \gamma}^{2\nu} \,
\frac12 \,
\sigma_{\alpha} \, \sigma_{\gamma}   
\left\{ \delta_{n_{\alpha}n_{\gamma}} 
A(r_{\alpha \gamma}) +
(1 - \delta_{n_{\alpha}n_{\gamma}}) \,
B(r_{\alpha \gamma}) \right\}
\right) \,,
\end{eqnarray}
where the contact terms $\alpha = \gamma$ are treated separately
(the latter modification leads 
to a physically irrelevant renormalization of the 
partition function). The frequency $b$ is inversely proportional 
to the temperature $b^2 = 2\pi/(k_{\rm B} T)$, and the Fourier 
amplitude $y$ is related to the fugacity $z$ by the relation 
$z^{2\nu}/(\nu!)^2 = (y/2)^{2\nu}/(2\nu)!$, i.e.~
$y = 2 z \left( \frac{(\nu + 1)_\nu}{\nu!}\right)^{1/(2 \nu)}$
where $(a)_n = \Gamma(a+n)/\Gamma(a)$ is the Pochhammer symbol.

%
%
\subsection{Comparison of the layered sine--Gordon and 
layered vortex gas models} 

The partition functions (\ref{LVG}) of the layered vortex gas
and (\ref{EGTE}) of the layered sine--Gordon gas
have the same structure. Therefore, the intra- and interlayer interaction 
potentials can thus be compared directly. 
A comparison of Eqs.~(\ref{int_pot_intra}) and (\ref{intpot_intra}) reveals 
that for small distances ($r\ll \lambda_{\mathrm{eff}}$), the intralayer 
potentials have the same logarithmic behaviour for both models. This is 
not unexpected since in this case the vortices of a given layer are 
independent of the effects in the other layer. However, for large distances 
($r\gg \lambda_{\mathrm{eff}}$), the intralayer potential is logarithmic for 
the gas of topological excitations of the layered sine--Gordon model in 
contrast to the the layered vortex gas, whereas the long-range intralayer 
potential is dominated by a linear term. 
The difference of the two models becomes even more significant if one 
compares the interlayer potentials which are different for the two 
models both in the short-range as well as the long-range regime 
(see Eqs.~(\ref{int_pot_inter})  and~(\ref{intpot_inter})). 

The significant differences of the long-range behaviour of the interlayer 
potentials strongly indicate different long-distance (infrared, IR) physics.
We should note that the long-range behaviour of the potentials in 
Eqs.~(\ref{intpot_intra_long}) and (\ref{intpot_inter_long}) generalizes 
to a leading asymptotics of the form 
$-1/(2 \pi N) \ln(r_{\alpha\gamma}/\lambda_{\mathrm{eff}})$ for an 
$N$-layer system with the interlayer interaction given in Eq.~(\ref{NLSG}) 
below. Thus, the addition of more layers thus does not change the qualitative 
behaviour of the long-range potentials (linear versus logarithmic).
We conclude that there is a strong indication that these models belong to 
different universality classes, and that the layered sine--Gordon model 
is {\em not} suitable to describe the vortex properties of Josephson 
coupled layered superconductors if a linear long-range potential between 
the topological defects is assumed. 

%
%
\section{PHASE STRUCTURE OF THE $\boldsymbol N$-LAYER SINE--GORDON MODEL}
\label{rg}

From a conceptual point of view, it is interesting to study 
the critical temperature as a function of the number of coupled layers,
in the framework of an appropriate generalization of the double-layer
model defined in Eq.~(\ref{2LSG}) to $N$ layers. 
The discretization of the derivative term for the $z$-direction in the 
three-dimensional sine--Gordon Lagrangian results in a model of coupled 
2D systems~\cite{Na2006}, which has been called the $N$-layer sine--Gordon 
model. It consists of $N$ coupled 2D sine--Gordon models of identical 
``frequency'' $b$~\cite{JeNaZJ2006,Na2006}, each of which corresponds to a 
single layer described by the scalar fields $\varphi_i$ $(i=1,2,\ldots,N)$.
Its bare action reads (see Eq.~(2) of Ref.~\cite{JeNaZJ2006})
\begin{align}
\label{NLSG}
S_{\mathrm{NLSG}}
=\int {\rm d}^2 r \left[ \hf \sum_{i=1}^N (\partial_{\mu}\varphi_i)^2 + 
\sum_{i = 1}^{N-1} \frac{J}{2} (\varphi_{i+1}-\varphi_i)^2 
+ \sum_{i=1}^N y_i \cos (b\varphi_i) \right]\,.
\end{align}
We have implicitly defined the mass 
matrix $\underline{\underline m}^2_N$ of the $N$-layer model, 
$\sum_{i = 1}^{N-1} \frac{J}{2} (\varphi_{i+1}-\varphi_i)^2
\equiv \hf \underline{\varphi}^{\rm T} \, \, \underline{\underline m}^2_N 
\, \, \underline{\varphi}$. The action is invariant under a joint 
shift of all fields $\varphi_i\to \varphi_i+ 2\pi/b$ applied to all 
layers $i=1,2,\ldots,N$, a symmetry which corresponds to a single zero 
mass eigenvalue of the matrix ${\underline{\underline {m}}}^2_N$. Indeed, 
after a suitable rotation of the mass matrix~\cite{JeNaZJ2006,Na2006}, 
it becomes evident that the $N$-layer sine--Gordon model consists 
of $N-1$ massive 2D and a single massless 2D sine--Gordon fields. The 
periodicity in the internal space spanned by the field 
is broken explicitly for the massive fields, and the 
spontaneous breaking of periodicity of the single massless mode 
accompanies the phase transition for small values of 
fugacities~\cite{NaNaSaJe2005,JeNaZJ2006}.

The rotated $N$-layer sine--Gordon model has already been investigated 
by the Wegner--Houghton renormalization group method on the basis of the 
mass-corrected linearized scaling laws \cite{Na2006} and by a general 
perturbative treatment~\cite{JeNaZJ2006}. Both approaches predict a 
linear increase of the critical parameter $b_c^2$ with increasing number 
$N$ of the layers, according to the formula, $b^2_{c} = 8 \pi N$,
(see Eq.~(35) of Ref.~\cite{JeNaZJ2006}).
Equation~(\ref{EGTE}) clearly implies 
$b^2_{\rm c} = 8 \pi N = 2 \pi/(k_{\rm B} T)$, and we therefore obtain
$T_{\rm c} \propto 1/N$
for the $N$-layer model. This decrease of the transition temperature 
is perfectly consistent with the general properties of the model in the 
limit of an infinite number of layers. Namely, one can intuitively assume 
that the single remaining zero-mass eigenvalue cannot make a decisive 
contribution to the phase structure of the model in the limit $N \to \infty$, 
with $N-1$ modes being massive. Indeed, in the limit of an infinite number of 
layers, one recovers the 3D sine--Gordon model which does not undergo
any phase transition at all~\cite{Ko1977,NaJeSaSo2004,Na2006}.

\section{SUMMARY AND CONCLUSIONS}
\label{sum} 

The main result of this paper (see Sec.~\ref{lsgsect}) is the indirect 
comparison of the layered sine--Gordon model to the layered Ginzburg--Landau
theory: as we have shown,
both models can be mapped to different gases 
of topological excitations. These are the layered vortex gas for the 
layered Ginzburg--Landau theory (see Eq.~(\ref{LVG})) and the equivalent 
gas of topological excitations for the layered sine--Gordon model 
(``layered sine--Gordon gas, see Eq.~(\ref{EGTE})). 
In general, we find that 
if a long-range confining linear potential is required for a 
description of the Josephson-coupled layered high-$T_{\rm c}$
superconductors,
then the system of coupled 2D sine--Gordon models is not suitable to 
describe the vortex properties of these materials: scalar-field 
propagators cannot provide linear potentials in two dimensions. For short 
distances, a logarithmic behaviour can of course be approximated quite well 
by a linear potential~(see Ref.~\cite{3DXY}, $\ln(1+r)\sim r$ for $r\ll 1$),
but this observation is 
irrelevant for the phase-structure of a system, which is determined only 
by the long-range interactions. In any case, we are led to the conjecture 
that the layered sine--Gordon and the layered vortex gas models belong to 
different universality classes. 
Using a renormalization group analysis of the
generalized $N$-layer sine--Gordon model as described in Sec.~\ref{rg},  
we find that the critical temperature of the 
layered sine--Gordon gas reads fulfills
$k_{\rm B}\, T_{\rm c} = (4 N)^{-1}$. This is 
inconsistent with high transition temperatures for multi-layer system 
and in strong disagreement with 
experiment~\cite{MaEtAl1992,MaEtAl1993,OtEtAl1994}. 
E.g., in Ref.~\cite{MaEtAl1993},
for $\mbox{YBa}_2 \mbox{Cu}_3 \mbox{O}_{7-\delta}$ 
the single-layer (2D) transition temperature
was determined as $30.1$~K, and with $N=2$ layers,
the experimental result was $T_{\mathrm{KTB}} = 58.2$~K,
suggesting $T_{\rm c} \propto N$ for a small number of layers.

Let us conclude this paper with a perhaps somewhat surprising outlook. 
The decrease of the transition temperature with the number of layers is 
tied to the gradual ``disappearance'' of the ``influence'' of the only 
remaining zero-mass mode in the matrix of the Josephson-coupled layered 
sine--Gordon model, in comparison to the $N-1$ massive modes, as 
$N \to \infty$. If we choose the mass matrix differently, e.g.,
$\underline{\varphi}^{\rm T} \, {\underline{\underline M}}^2 
\underline{\varphi} = G \left(\sum_{n=1}^N a_n \varphi_n \right)^2$,
with the (only) condition $a_n^2 = 1$, then there are $N-1$ massless 
modes and only one massive mode. In that case, we find 
(see~\cite{NaEtAl2007prb}), $T_{\rm c} \propto \frac{N-1}{N}$,
and this result is in agreement with the analysis presented in
Ref.~\cite{CoGeBl2005} for magnetically coupled layered superconductors. 
In this case, the interaction potentials corresponding to 
Eq.~(\ref{intpot_limits}) between the topological defects have the 
same asymptotic behaviour as those given in 
Refs.~\cite{BlEtAl1994,CoGeBl2005} for the magnetically
coupled case. After all, a layered sine--Gordon type field theory with a 
suitable interlayer interaction might prove to be useful for the 
description vortex dynamics in (magnetically coupled) layered systems, 
but not in the expected direction, which would 
have been the Josephson-coupled case.

\end{document}